\documentclass[
aps,%
11pt,%
final,%
notitlepage,%
oneside,%
twocolumn,%
nobibnotes,%
nofootinbib,%
superscriptaddress,%
noshowpacs,%
centertags]%
{revtex4}

\begin{document}

\title{ANALYSIS OF THE RC CATALOG SAMPLE IN THE REGION OVERLAPPING WITH
THE REGIONS OF THE FIRST AND SDSS SURVEYS.
I. IDENTIFICATION OF SOURCES WITH THE VLSS, TXS, NVSS, FIRST, AND GB6
CATALOGS}
\author{\firstname{O.~P.}~\surname{Zhelenkova}}
\affiliation{\saoname}
\author{\firstname{A.~I.}~\surname{Kopylov}}
\affiliation{\saoname}

\begin{abstract}
Radio sources of the RC catalog produced in 1980--1985 at \mbox{RATAN-600}
radio telescope based on a deep survey of a sky strip centered on the
declination of the SS~433 source are optically identified in the region
overlapping with FIRST and SDSS surveys (about 132\,$\square\degr$).
The NVSS catalog was used as the reference catalog for refining the
coordinates of the radio sources. The morphology is found for about 75\% of
the objects of the sample and the ratio of single, double and multicomponent
radio sources is computed based on FIRST radio maps. The 74, 365, 1400, and
4850~MHz data of the VLSS, TXS, NVSS, FIRST, and GB6 catalogs are used to
analyze the shape of the spectra.
\end{abstract}

\maketitle

\section{INTRODUCTION}

The approaches toward the panchomatic study of extragalactic sources including radio
sources  based on statistical analysis of the properties of large samples
produced using automated procedures of data reduction have become
increasingly important in late 20--early 21 century.
Identification of sources at radio frequencies is a  difficult task because
of different angular resolutions, coordinate accuracy, and limiting
sensitivity of the radio catalogs compared.
The procedure must also take into account the flux density variation of the
source at
different frequencies. However, surveys with high precision of coordinate
measurement  (on the order of one arcsec)---such as NVSS \cite{Condon} and
FIRST \cite{Becker} have been subject to mass cross identification with
optical surveys.

Thus, e.g., McMahon et al.~\mbox{\cite{McMahon1}} identified a total of
382892 radio sources of the  FIRST survey ($S_{1.4\ GHz}^{lim}$>1\,mJy)
with the APM survey~\cite{McMahon2} within a 4150\,$\square\degr$ large
region in the vicinity of North Galactic Pole. To identify multicomponent
sources, the cross-identification algorithm employed allowed for an
empirical relation between flux and separation between the components and
assigned accordingly the probability of associating the objects located
within the search region with a single radio source. The adopted maximum
component separation of \mbox{$r<120\arcsec$} implied that about 8\%, 3\%,
and about 1\% of the sources are double, triple, or consist of four or more
components, respectively.
With these results taken into account, the total fraction of identifications
down to the limiting magnitude of E=$20.5^m$ of the POSS-I survey
\cite{Abell} amounted to about 24\%.

Ivezi$\acute c$  et al.~\cite{Ivezic} analyzed the radio and optical
properties of about 30000  FIRST radio sources whose coordinates coincided
with those of  ERD SDSS objects \mbox{($r^*\sim22.2^m$) \cite{Stoungton}}.
They automatically cross identified the two catalogs with a circular search
region of radius $1.5\arcsec$ centered on the position of the optical object.
The above authors analyzed only point sources during the first pass. During
the second pass they selected from among the FIRST sources that were not
identified with optical objects the pairs of neighboring sources (with
interpair separations $r<90\arcsec$) and performed the second cross
identification using the position of the midpoint between the components.
Radio sources with more complex morphology (i.e., non-point sources) proved
to make up for less than 10\% of the total number of objects of the FIRST
catalog. The fraction of optically identified radio sources was found to be about 27\%.

In 1980--1985 a deep multifrequency survey of a $20\arcmin$-wide sky strip
centered on the declination of SS433 ($\delta_{1950.0} = +4\degr54\arcmin$)
was performed with the \mbox{RATAN-600} radio telescope. The angular
resolution of the survey was equal to $\Delta\alpha\sim\,1\arcmin$ for
$\lambda=7.6$\,cm \mbox{(3.9\,GHz) \cite{Berlin1, Berlin2}.} The
observational data named ``Cold'' were used to produce the  RC catalog, which includes 1165
sources from \mbox{\cite{Parijskij} and \cite{Par}.} The version of the
catalog stored in the CATS database \cite{Verkhodanov}, which we used,
contains a total of  1209 objects.

The recent release of a number of radio surveys---namely,
{VLSS \cite{Cohen, Cohen1}}, TXS \cite{Douglas}, and GB6
\cite{Gregory}---which covered the strip of the ``Cold'' survey, made it
possible to study  the properties of the sources of the RC catalog based on
the 74\,MHz, 365\,MHz, 1.4\,GHz, and 4.85\,GHz data of the surveys. To refine
the coordinates for the sample of the radio sources of the RC catalog and
obtain information on the morphology of radio sources for the subsequent
optical identification with the \mbox{SDSS \cite{Adelman}} survey, we
selected radio surveys with high angular resolution and high coordinate
accuracy---NVSS and FIRST. We planned to identify all sources of the RC
catalog located within the area overlapping with the regions covered by
the FIRST and SDSS surveys, namely, the strip with the area of about
132\,$\square\degr$ located from $\alpha_{2000.0}$ = $8^h11^m$ to
$\alpha_{2000.0}$ = $16^h25^m$ and from $\delta_{2000.0}$ = $+4\degr20\arcmin$
to $+5\degr24\arcmin$. No detailed analyses of radio sources with flux densities
from $S_{3.9\,GHz}\sim 11$~mJy and higher have been performed so far, and
such studied may be of certain interest, because we imposed no additional
constraints on morphological type, spectral index, or angular size.

The beam pattern of the \mbox{RATAN-600} has a ``knife-edge'' shape at the
elevation of ``Cold'' survey observations, $H=51\degr$. The declination
angular resolution of the telescope in about three times lower than the
resolution in right ascension \cite{Esepkina, Majorova}. The widely used
\mbox{ConeSearch \cite{Simple}} algorithm of automatic catalog cross
identification uses the same search radius for both coordinates. The more
sophisticated \mbox{SPECFIND \cite{Vollmer}} algorithm, which
takes into account the angular resolution of the catalogs compared and
spectral features of the sources when identifying the sources assumes that
the beam pattern of the telescope has identical resolution in both
coordinates and that coordinate errors are smaller than the error of angular
resolution. These  algorithms yield low percentage of coincidences for cross-matching of
the RC catalog and
that is why identification was performed by visually inspecting superposed
optical and radio images and analyzing the data from the catalogs and
surveys selected for this work. The data for each source were prepared
automatically using a Perl script code for the interface of  Aladin
interactive sky atlas \cite{Ochsenbein1}.

In this paper we describe the technique of identification and report the
results of identification of the RC catalog with the  VLSS, TXS, NVSS,
FIRST, and GB6 surveys. We use the refined coordinates and morphology of
radio sources to perform optical identification of the sample and we plan to
report the results in a separate paper. When operating with heterogeneous
data of catalogs and surveys we used virtual observatory software tools
Aladin, Vizier \cite{Ochsenbein2} and TOPCAT \cite{Taylor}.
	
\section{TECHNIQUE OF IDENTIFICATION}

The accuracy of the definition of the coordinates of the sources observed with
RATAN-600 depends on their relative elevation $\Delta H$ with respect to the
center of the  beam pattern of the telescope  and on flux densities
\cite{Soboleva, Majorova}.
For our sample the median errors of RC source
coordinates as given in the version provided by CATS data base are equal to
0.58$^{s}$ and $38.0\arcsec$ in right ascension and declination, respectively.
For optical identification the error of the coordinates of the radio sources
must not be greater than 1 arcsec. We refined the coordinates of the radio
sources using the NVSS and FIRST surveys.
The angular resolution of the RC catalog in right ascension is close to that
of the NVSS survey and therefore we first analyzed the position of the source
in the RC catalog with respect to NVSS. Due to its high angular resolution
the FIRST survey provides detailed information about the structure of the
source, which is required for optical identifications, but makes it difficult
to identify the object at radio wavelengths.

For each source of our list we saved the results of queries to the selected
data resources  in the Aladin stack. To visualize the mutual arrangement of
radio sources in the stack we overlaped the isophotes of NVSS and FIRST
images and indicated the positions of the objects of the selected radio catalogs.

Below we list the conditions (in the order of decreasing importance) that we
took into account when identifying an RC source with an reference-catalog
object.
\begin{itemize}
\item Coordinate agreement in right ascension---the separation between the
RC radio source and the corresponding object of the NVSS or FIRST catalog
must satisfy the inequality $r<3\sigma$, where $\sigma$ is the
right-ascension error quoted in the  RC.
\item Coordinate agreement in declination---similar to the agreement
in $\alpha_{2000.0}$: the separation between the RC position and the NVSS
or FIRST position must not exceed $r<3\sigma$.
\item Agreement between the flux density of the object as indicated in the catalog
studied and in the reference catalog. The cases rase certain doubts where
the source coordinates in the two catalogs agree with each other, but the
flux density does not agree with the  NVSS flux (we convert flux density assuming that the
spectral index of the source is $\alpha\sim 0.7$,\ \mbox{$S(\nu)\sim
\nu^{-\alpha}$).}
\item Presence of neighboring sources. If an RC source has two or more
reference-catalog sources located within the beam pattern of
RATAN-600, such sources blend together, making identification ambiguous.
In these cases we assumed that the  brightest source provided the greatest
contribution and identified it with the RC object considered.
\end{itemize}
We also took into consideration the possible effect on coordinate
measurements and inferred fluxes due to bright objects located at separations
exceeding the size of the beam pattern. A group of faint objects
within the beam pattern may have a similar effect.

To take this effect from the neighboring sources into account and resolve
the ambiguities in identification, we use an atlas of the ``Cold'' survey
strip. The figures with  areas with the size of 15 minutes in right ascension
and $1\degr 30\arcmin$ in declination show the RC sources with the
corresponding error bars ($3\sigma$) as well as the positions and flux densities of
the radio sources from the VLSS, NVSS, TXS, GB6, PMN, and a number of other
catalogs~\cite{Kopylov}.

In the cases where the 3.9 GHz flux density of a source exceeds the  1.4~GHz one
given in the reference catalog---additional information is required to
confirm the increase of flux density toward higher frequencies. In these cases we
use both the catalog and the GB6 survey. The catalog usually includes
objects with flux densities greater than $5\sigma$ of the signal-to-noise level.
Sources with flux densities at the $3\sigma$--$4\sigma$ level, which are absent in
the  GB6 catalog, can be found by visually inspecting the images of the  GB6
sources. This additional information was of great assistance when we dealt
with cases of ambiguous identification.

After examining the stacks and atlas and comparing the data from selected
radio catalogs we subdivide RC sources into three groups as follows:
\begin{itemize}
\item
``RC''---the source can be confidently identified not only with an object in
the NVSS and a FIRST catalogs, but also with an object in  at least one of
the following catalogs: VLSS, TXS, or GB6;
\item
``rc''---the source can be identified with an object of the  NVSS and/or
FIRST survey;
\item
``X''---the source could not be unambiguously identified with any object
from other catalogs.
\end{itemize}

Of the 432 RC radio sources located within the region of intersection with
the SDSS and FIRST surveys, we classified 190 (44\%), 130 (30\%), 98 (23\%),
and 14 (3\%) objects as belonging to the ``RC'',``rc'', ``X'', and  ``twin''
- object groups, respectively. The RC catalog contains objects marked
as ``t'' (twin), which means that for the source in question there are
different variants of identification to interpret the observed scans in the
meridian and azimuth and, consequently, at least two variants of coordinates
\cite{Parijskij}.  Hereafter we analyze 320 radio sources belonging to
the  ``RC'' and ``rc'' groups.

\section{ANGULAR SIZES, THE NUMBER OF COMPONENTS, AND STRUCTURE OF THE
SOURCES}

For RC objects classified as belonging to the ``RC'' and  ``rc'' groups we
found the angular sizes, number of components, and morphological structure
according to the data of the  FIRST survey. To combine FIRST objects into a
single source or, on the contrary, to treat them as independent sources, we
analyze for each RC radio source contour map drawn using the service that
constructs isophotes of radio images for the FIRST survey without loss
of angular resolution \cite{Service}.

We estimate the angular sizes of RC sources depending on their morphological
structure as inferred from isophotes in the FIRST survey. We set the angular
sizes of single-component RC sources equal to the major axis of the
corresponding  FIRST catalog source. We use Aladin tools to determine the
angular sizes of multicomponent sources as the angular distance between the
farthermost components.

Table~\ref{ncomp} lists the angular sizes and the fractions of
single-component and multicomponent sources. Our counts include 318 of 320
sources (two faint extended NVSS sources are absent in the FIRST survey).
Note that some of the single-component sources can be resolved and have
nonpoint structure.

\begin{table*}[tbp]
\setcaptionmargin{0mm} \onelinecaptionstrue
\captionstyle{flushleft}
\caption{The fraction of single-component and multicomponent sources and the
angular sizes (the Largest Angular Size or LAS) of 318 RC objects (according
to the data of the  FIRST survey)}
\label{ncomp}
\bigskip
\begin{tabular}{c|c|c|c}
\hline
Number of components & Fraction of the sample & Number of objects & $LAS_{median}$ \\
          & (\%)           &                & ($\arcsec$)     \\
\hline
1       & 56  & 177  & 1.83  \\
2       & 27  & 85   & 17.5  \\
3       & 11  & 35   & 33.1  \\
4       & 4   & 14   & 60    \\
$\ge 5$ & 2   & 7    & 94    \\
\hline
\end{tabular}
\end{table*}

By the results of automatic cross identification between FIRST and
SDSS surveys Ivezic et al.~\cite{Ivezic} found that 90\%
of all radio sources to consist of a single component, whereas 10\% of the
sources have several components.
McMahon et al.~\cite{McMahon1} report a similar result (about 12\% fraction
of multicomponent sources) according to automatic identification of FIRST
and APM surveys, although Cress et al.~\cite{Cress} believe this fraction to
be higher and estimate it at about 16/
objects. The authors of~\cite{Cress} assume that if the angular separation
between the objects of the catalog does not exceed  $0.02\degr$, the
corresponding objects are components of the same source. We found the
fraction  single-component sources in our sample to be  28\% less than the
estimate reported by Cress et al.~\cite{Cress} and one and a half times more
than the estimates reported by McMahon et al.~\cite{McMahon1} and Ivezic
et al.~\cite{Ivezic}. A comparison of the number of objects in the FIRST
catalog with the number of real radio sources yields a ratio of 5:3 for our
sample, i.e., there are three radio sources for every five objects.

Lawrence et al.~\cite{Lawrence} give the morphological classification of the
radio sources of the MIT-Green Bank (MG) survey based on 4885\,MHz VLA maps
with the angular resolution of $0.4\arcsec$ or $1.2\arcsec$, which includes
10 types, namely:
\begin{enumerate}
\item[1)]
point---a point radio source that cannot be resolved into components;
\item[2)]
quasi-point---a radio source dominated by a point core with inconspicuous
structure;
\item[3)]
diffuse---a resolved source with poorly discernible intensity peaks;
\item[4)]
core-jet---an unresolved peak with an extension on one side or with a
closely located faint extended component;
\item[5)]
cometary---similar to a type 4 source, but with a resolvable peak;
\item[6)]
double---a source with two approximately symmetric components of about the
same flux density;
\item[7)]
triple---a triple source;
\item[8)]
multiple---four or more well-defined peaks;
\item[9)]
core-double---unlike type 7 sources, these objects have a fainter core and
extended components;
\item[10)]
jet---two relatively symmetric jets, sometimes with a discernible core and
with no other compact regions.
\end{enumerate}

\begin{table*}[tbp]
\setcaptionmargin{0mm} \onelinecaptionstrue
\captionstyle{flushleft}
\caption{Distribution of morphological types of the 320 RC sources}
\label{morph}
\bigskip
\begin{tabular}{l|c|c|c}
\hline
Type & Fraction of the sample & Number of objects & $LAS_{median}$ \\
    & (\%)           &                & $(\arcsec)$ \\
\hline
C (core)           & 39      & 125      & 1.42  \\
CL (core-lobe)     & 6       & 18       & 12.5  \\
CJ (core-jet)      & 8       & 24       & 7.28  \\
D (double)         & 33      & 106      & 13.6  \\
DC (core-double)   & 6       & 19       & 49.8  \\
DD (double-double) & 1       & 4        & 60.3  \\
T (triple)         & 6       & 20       & 34.9  \\
M (multiple)       & 0.5     & 2        & 33.1  \\
E (diffuse)        & 0.5     & 2        & --    \\
\hline
\end{tabular}
\end{table*}

We used for our classification the images of the  FIRST survey whose angular
resolution of $5.4\arcsec$ is close enough to that of the MG-VLA
\cite{Lawrence} survey and therefore we based our classification on the
above schema somewhat modifying it in order to reflect the relation between
the structure of the radio source and the position of the host galaxy, which
shows up clearly in many cases. We identified the following morphological
types:
\begin{enumerate}
\item[1)]
core (C)---a point radio source, which cannot be resolved into components
(this type includes the point and quasi-point types of the classification
above). The optical object coincides with the radio intensity peak;
\item[2)]
core-jet (CJ)---an unresolvable peak extending on one side or having a
nearby faint extended component (this type includes  core-jet and cometary
types of the classification above); the optical object coincides with the
maximum of radio intensity peak;
\item[3)]
core-lobe (CL)---a radio source with a core and components whose brightness
decreases toward the periphery (includes the  jet type); the optical object
coincides with the maximum of radio flux density;
\item[4)]
double (D)---a two-component radio source. The brightness of the components
increases toward the periphery of the source (FRII), the optical source is
located between the components. Double sources include two more types:
\begin{itemize}
\item
core-double (DC)---a two-component source with a faint core, the optical
object coincided with the core;
\item
double-double (DD)---a source similar to DC, but with the components
exhibiting a well-defined double structure,
the optical object is located between the radio components;
\end{itemize}

\item[5)]
triple (T)---a triple source. The central component resembles a point source
and the optical object coincides with the central component;
\item[6)]
multiple (M)---a multicomponent source whose structure that does not match
any of the above cases; additional information is required for optical
identification;
\item[7)]
diffuse or extended (E)---an extended source (may be absent in FIRST,
although present in NVSS); additional information is required for optical
identification.
\end{enumerate}

Table~\ref{morph} gives the distribution of morphological types of 320 RC
radio sources and the median angular size for each group. Note that when
assigning a certain type to the source we examined  NVSS and FIRST images
and took into account the position of the likely candidate for identification.

\begin{figure*}[tbp]
\setcaptionmargin{5mm}
\onelinecaptionsfalse
\includegraphics[scale=0.6]{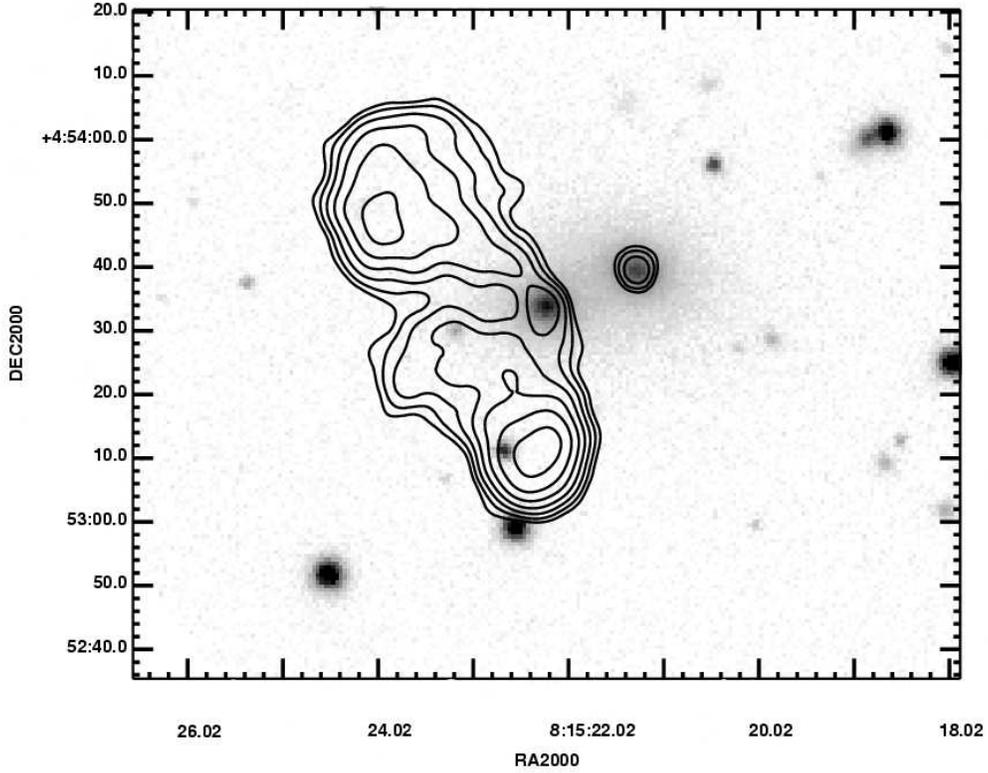}
\captionstyle{normal}
\caption{
RC J0815+0453. A group of two radio sources. Each source is identified with
an elliptical galaxy. The isophotes from the FIRST survey are superimposed on the
corresponding r-filter image of the SDSS survey.
}\label{RC0815}
\end{figure*}

Radio sources with asymmetric structure---the so-called ``winged'' or
``X-shaped'' sources---constitute a small and interesting population of radio
galaxies~\cite{Cheung}. In addition to the common pair of radio components,
these objects have a pair of low surface brightness emitting regions, which
form the wings or have an X-like shape. This shape is believed to be a result
of plasma outflow from hot-spot regions into the inhomogeneous medium
surrounding the radio source~\cite{Leahy}. Another explanation associates
the unusual structure of the source with the fact that low-brightness regions
may be residual phenomena of the rapid change of the rotational orientation
of the system with a supermassive black hole (SMBH) and
an accretion disk that occurred because of a recent merger of a double
SMBH \cite{Dennett}. X-shaped sources are of interest as systems associated
with double black holes \cite{Komossa} and recurrent
phases of the activity of the radio source in the host
galaxy \cite{Liu}. We computed the number of radio sources with a close
to  ``winged'' or ``X-shaped'' structure. Such sources make up for
about 4\% of all the sources considered.

There are groups consisting of several sources and there are pairs located
within  1-1.5 arcmin of each other. We found them to make up for about
7\% of all the cases. Figure~\ref{RC0815} shows an interesting example
of a pair consisting of a double and a faint point radio source identified
with two elliptical galaxies.

The results of the identification of radio sources with the VLSS, TXS, NVSS,
FIRST, and GB6 surveys, the angular sizes, morphology, and spectral indices
can be found in the electronic table available at
{\tt {\small http://www.sao.ru/hq/zhe/RCriResInn.html}}. One can also find
there a description of the columns and
contour maps of RC radio sources based on images of the FIRST survey.

\section{THE 74--4850\,MHZ SPECTRA OF RADIO SOURCES}

\begin{figure*}[tbp]
\setcaptionmargin{5mm}
\onelinecaptionsfalse
\includegraphics[scale=0.8]{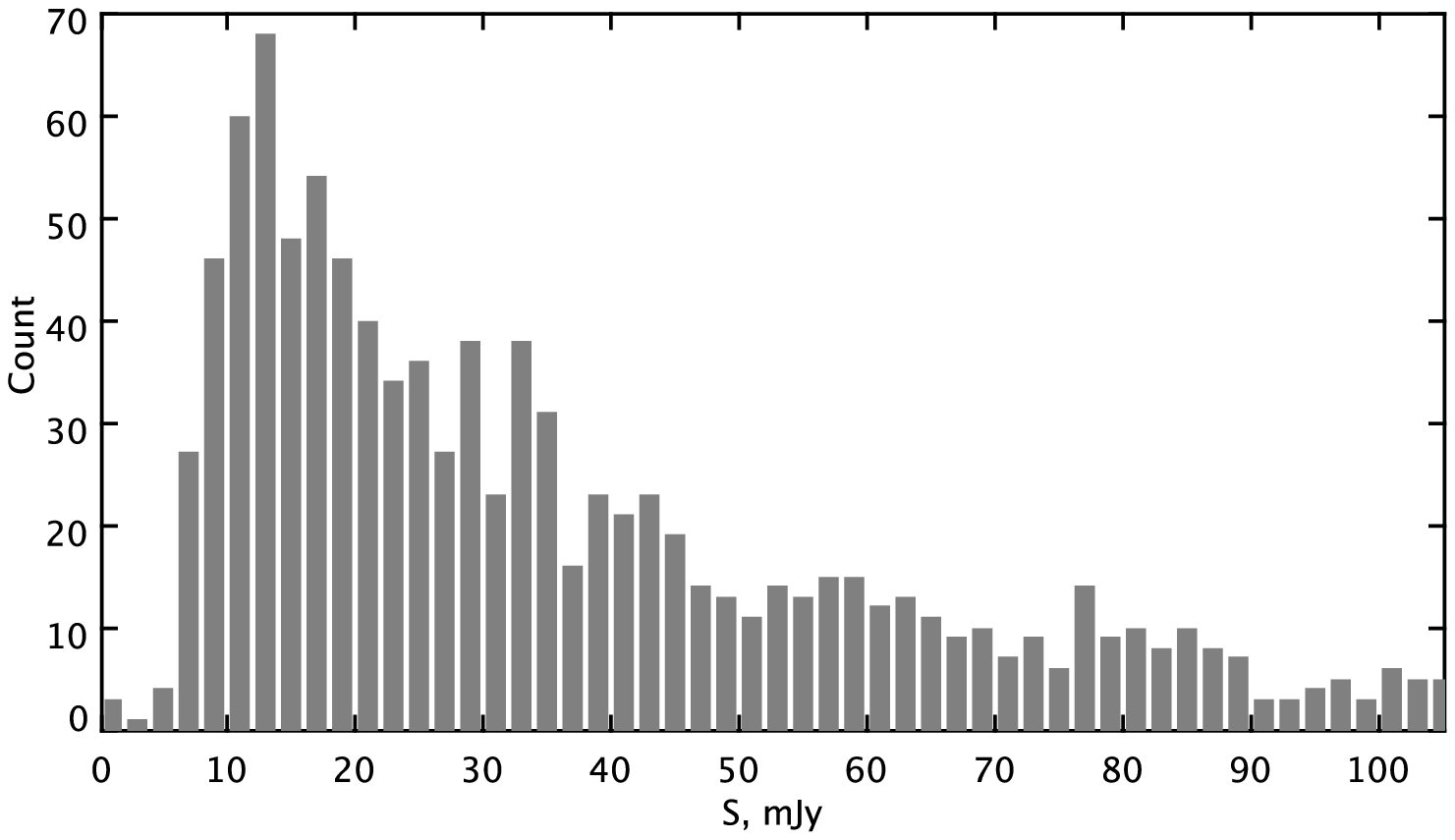}
\captionstyle{normal}
\caption{
   Distribution of the 3.9\,GHz flux densities of the RC catalog sources
(sources brighter than 105\,mJy are not plotted). The number of sources
decreases sharply at flux densities <12\,mJy.
   }\label{diffn}
\end{figure*}

The completeness of the RC catalog in the central part of the  ``Cold''
survey is close to unity for radio sources with flux densities
\mbox{$S_{3.9\,GHz}$>~15\,mJy \cite{Soboleva}.} This region includes the
$10\arcmin$ wide declination strip centered on the declination of
the  SS\,433 source at the epoch of the survey.
Figure~\ref{diffn} shows the distribution of the flux densities of radio sources.
The sharp decrease for faint sources starts at $S_{3.9\,GHz}$<~11--12\,mJy.
The catalog lists all objects with flux densities $S_{3.9\,GHz}\ge 30$\,mJy in the
20$'$ strip without exception. When composing two flux limited samples we
took into account the declination difference between  SS\,433 and the
radio-source coordinates refined using NVSS and precessed for the epoch of
observations (15.04.1980), as well as the flux density refined using the corrected
$\Delta H$ and the computed directional diagram of the telescope for the
elevation of $H=51\degr$. The first complete sample includes sources with
elevation differed from that of the center of the directional diagram by no
more than $\Delta H\le |5\arcmin|$ and with fluxes equal to or greater
than  $S_{3.9\,GHz}\ge$11\,mJy (a total of 130 objects). The sample covers
an area of about 21\,$\square\degr$ The second sample consists of the sources
with $\Delta H\le |10\arcmin|$ and $S_{3.9\,GHz}\ge$~30\,mJy
(a total of 117 objects covering an area of about  41\,$\square\degr$).
The samples partly overlap, because  47\% of the sources of the second
sample belong to the first sample. Hereafter for the sake of brevity we
refer to the first and second samples as  1S and 2S, respectively.

\begin{table*}[tbp]
\setcaptionmargin{5mm} \onelinecaptionstrue
\captionstyle{flushleft}
\caption{Distribution of the types of the spectra  (of the spectral index
$\alpha_{1.4-4.85\;GHz}$) of the sources
in the 1S ($\Delta H\le |5\arcmin|$; $S_{3.9\;GHz}\ge$11\,mJy) and
2S ($\Delta H\le |10\arcmin|$; $S_{3.9\;GHz}\ge$30\,mJy) samples
}
\label{spind}
\bigskip
\begin{tabular}{c|c|c|c|c|c|c|c|c}
\hline
Spectrum & Sample & Fraction of the sample  & $N_{obj}$ & $LAS_{median}$  & $S_{1.4\;GHz}$ & $S_{3.9\;GHz}$ & $S_{4.85\;GHz}$ & $\alpha_{median}$ \\
       &         & (\%)            &           & ($\arcsec$)     & (mJy)          & (mJy)          & (mJy)           &          \\
\hline
I      & 1S      & 10 & 13        & 2.33          & 12.1  & 14 & 20 & -0.32 \\
       & 2S      & 6  & 7         & 2.33          & 34.3  & 68 & 51 & -0.32 \\
\hline
F      & 1S      & 27 & 35        & 2.34          & 39.1  & 22 & 28 & 0.31   \\
       & 2S      & 22 & 26        & 2.27          & 74.2  & 45 & 46 & 0.25  \\
\hline
S      & 1S      & 54 & 70        & 10.6          & 82.3  & 31 & 33 & 0.76  \\
       & 2S      & 56 & 65        & 10.26         & 146.5 & 57 & 59 & 0.76  \\
\hline
U      & 1S      & 9  & 12        & 17.1          & 94.5  & 31 & 22 & 1.17  \\
       & 2S      & 16 & 19        & 17.1          & 146.1 & 69 & 39 & 1.12   \\
\hline
\end{tabular}
\end{table*}

We used the images of the  GB6 survey to estimate the 4.85\,GHz fluxes for
the sources of the sample that were absent in the  GB6 catalog. For this, we
compared the peak intensities in the region of the radio source
studied and of the neighboring sources listed in the  GB6 catalog. Note that
for the GB6 catalog the 5$\sigma$ detection limit for radio sources located
at \mbox{$\delta\sim+5\degr$} is about 28\,mJy in the right-ascension
intervals \mbox{$8^{h}< \alpha_{2000.0}<12^{h}$} and
\mbox{$14^{h}40^{m}< \alpha_{2000.0}<16^{h}30^{m}$} and
about 37\,mJy in the right-ascension interval
\mbox{$12^{h}<\alpha_{2000.0}<14^{h}40^{m}$}~\cite{Gregory}. Out from the
130 sources of the 1S sample 51 source was identified with the GB6
catalog and for the remaining 79 sources we give the flux density estimates inferred
from the images of the  GB6 survey. For the 2S sample the corresponding
numbers are equal to 85 and 32, respectively.

We subdivided all sources into four groups in accordance to their spectral
index $\alpha_{1.4-4.85\,GHz}$ (see Table~\ref{spind}):
\begin{itemize}
\item
inverse (I), $\alpha<-0.1$;
\item
flat (F), $-0.1\le\alpha<0.5$;
\item
steep (S), $0.5\le\alpha<1$;
\item
ultrasteep (U), $\alpha\ge 1$.
\end{itemize}

\begin{figure*}[tbp]
\setcaptionmargin{5mm}
\onelinecaptionsfalse
\includegraphics[scale=0.4]{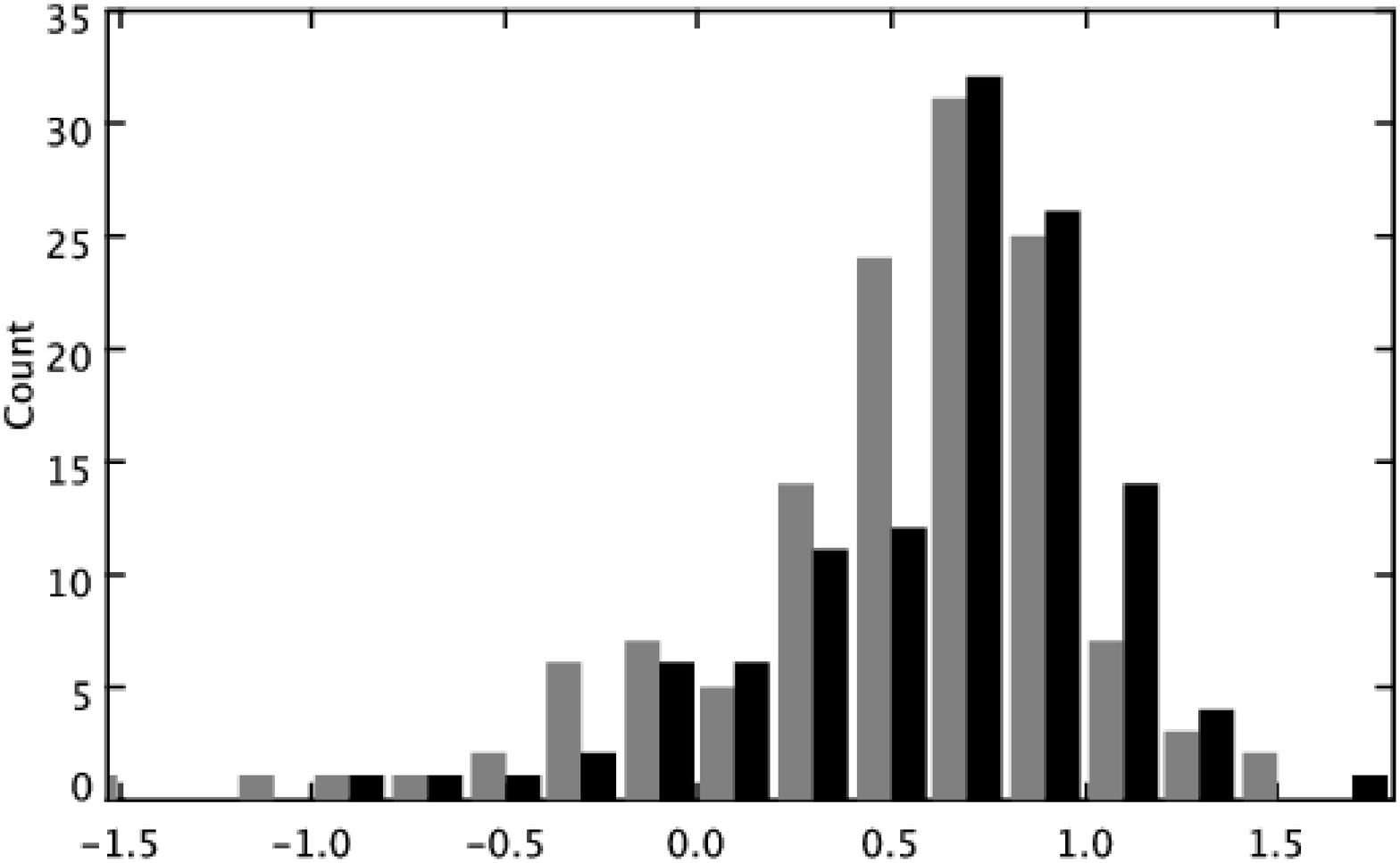}
\captionstyle{normal}
\caption{
   Distribution of spectral indices $\alpha_{1.4\,GHz-4.85\,GHz}$ for
sources from the central part of the ``Cold'' survey. The gray and black
histograms correspond to the sources of the 1S ($\Delta H\le |5\arcmin|$;
$S_{3.9\;GHz}\ge$11\,mJy) and 2S ($\Delta H\le |10\arcmin|$;
$S_{3.9\;GHz}\ge$30\,mJy) samples, respectively. The 1S sample, which
includes sources with smaller  3.9\,GHz flux densities compared to the
2S sample contains more flat and inverse spectra than the 2S sample.
   }\label{NG2samp}
\end{figure*}

The median flux density of the sources is equal to
\linebreak $S^{median}_{3.9\,GHz}$=26\,mJy and
$S^{median}_{3.9\,GHz}$=57\,mJy for the first (1S) and second (2S) samples,
respectively. A comparison of the spectral indices of the radio sources
shows that the number of sources with inverse and flat spectra slightly
increases and that of the sources with steep and ultrasteep spectra
decreases  (see Table~\ref{spind} and Fig.~\ref{NG2samp}).

The sources with flat and inverse spectra of the 1S and 2S samples proved to
be more compact in terms of angular sizes (see Table~\ref{spind}) than
sources with steep and ultrasteep spectra.

For the 143 sources identified in three or four catalogs (VLSS, TXS, NVSS,
and GB6)  additional flux density information is available, which allows the
behavior of their spectra to be studied in the frequency interval
74--4850~MHz.
We estimated the flux densities of objects with no data given in the VLSS or GB6
catalog by comparing the peak value in the region of the radio source with
the peak values for the neighboring sources listed in the corresponding
catalogs. We set the flux densities of the sources absent in the TXS catalog equal
to the 150-mJy sensitivity limit of the catalog. We then computed the
two-frequency spectral indices  $\alpha_{74-365\;MHz}$,
$\alpha_{365\;MHz-1.4\;GHz}$, and $\alpha_{1.4-4.85\;GHz}$. We found the
fractions of sources with inverse, flat, steep, and ultrasteep spectral
indices $\alpha_{74-365\;MHz}$ proved to be equal to  10\%, 38\%, 46\%, and
6\%, respectively.The sources of the first group proved to be more compact
($LAS_{median}=1.02\arcsec$) than other sources.

In each group we analyzed how the form of the spectrum varies from low to
higher frequencies by fitting the spectrum to a parabola using the {\it spg}
procedure of the standard FADPS reduction system of the first feeder
of \mbox{RATAN-600} \cite{Verkhodanov1}. We use symbols I, F, S, and U to
denote the corresponding regions of the spectrum as we pass from one portion
to another. We obtained the following results (see Table~\ref{specshape}):
\begin{itemize}
\item[1)]
Sources with $\alpha_{74-365\;MHz}$<-0.1 have two types of spectra:
one group has a steep spectrum at 1.4\;GHz, which becomes steep (IFS) or
ultrasteep (IFU) by 4.85\;GHz, and the second group has ultrasteep spectra
at 1.4\;GHz.
\item[2)]
Sources with flat spectra can be subdivided into the following three groups:
\item
The spectra of most of the sources (29\% of all sources) become steeper
with increasing frequency (FS, FU);
\item
A small number of sources maintain flat spectrum in the frequency interval
considered (F) or become inverse at high frequencies (FI);
\item
The third group includes sources with spectra
$\alpha_{365\;MHz-1.4\;GHz}$>0.5, and flatten out by  4.85\;GHz (FSF).
\item[3)]
Sources with steep spectra can also subdivided into three groups:
\item
Most of the sources (34\%) have steep spectra whose spectral indices remain
unchanged  ($\alpha>0.5$) as one passes from one spectral interval to
another. We fit Sc SU to a parabola. At high frequencies the spectra of SU
sources become ultrasteep and those of Sc sources remain steep);
\item
A small number of sources with flattened or inverse spectra at
4.85\;GHz (SF---3.5\%, SI---1.5\%);
\item
The third group---transition from a steep to a flat spectrum, and then back
to a steep spectrum (SFS).
\item[4)]
The small group of sources with ultrasteep spectra in the frequency
interval 74--365\;MHz can be subdivided into the following two sub groups:
\item
one of the subgroups maintains its steep spectrum at higher frequencies,
but its slope is smaller than the initial slope (US);
\item
the spectra of the objects of the second subgroup becomes flat (UF) or even
inverse (UFI).
\end{itemize}

\begin{table*}[tbp]
\setcaptionmargin{0mm} \onelinecaptionstrue
\captionstyle{flushleft}
\caption{Variation of the spectral indices  $\alpha_{74-365\;MHz}$,
$\alpha_{365\;MHz-1.4\;GHz}$, and $\alpha_{1.4-4.85\;GHz}$ for 143
RC sources according to the data of the VLSS, TXS, NVSS, and GB6 surveys
}
\label{specshape}
\bigskip
\begin{tabular}{l|c|c|c|c}
\hline
$\alpha_{74-365\;MHz}$ & Form of the spectrum  & Fraction in the sample &Number of objects & $LAS_{median}$ \\
               &                & (\%)           &               & ($\arcsec$)   \\
\hline
I         & IFS & 3    & 4  & 0.91 \\
15 (10\%) & IFU & 2    & 3  & 1.35 \\
      & IU  & 5    & 8  & 1.02 \\
\hline
      & F   & 4    & 6  & 11.0 \\
F         & FS  & 15   & 22 & 5.25 \\
55 (38\%) & FU  & 14   & 20 & 7.02 \\
      & FI  & 1.5  & 2  & 1.40 \\
      & FSF & 3.5  & 5  & 4.83 \\
\hline
      & S   & 11   & 15 & 13.6 \\
S         & Sc  & 13   & 19 & 20.9 \\
65 (46\%) & SU  & 10   & 14 & 14.0 \\
      & SF  & 3.5  & 5  & 4.57 \\
      & SI  & 1.5  & 2  & 5.9  \\
      & SFS & 7    & 10 & 51.6 \\
\hline
U         & US  & 2    & 3  & 53.0 \\
8 (6\%)   & UF  & 3    & 4  & 16.5 \\
      & UFI & 1    & 1  & 0.76 \\
\hline
\end{tabular}
\end{table*}

\section{CONCLUSIONS}

We cross identified 432 radio sources of the RC catalog located within the
region of intersection of the SDSS and FIRST surveys of the FIRST, NVSS,
TXS, VLSS, and GB6 catalogs. In our statistical study and optical
identifications we use the sources (about 75\% of all sources) identified
with the NVSS and/or FIRST catalogs.
Most of the remaining RC objects (about 25\% of all sources) are either
false or blends of two or more real objects whose individual properties are
difficult to determine from observations of the ``Cold'' survey.

We use the data of the  FIRST survey to compute the number of components
(entries of the FIRST catalog) for the sources studied. We found that
single-component sources and sources with two or more components account for
about  56\% and more than 44\% of the entire sample, respectively.

This result is inconsistent with the results of automatic cross
identification between the  FIRST survey on the one hand and SDSS and
APM surveys on the other hand~\cite{McMahon1, Ivezic}, and the estimate  (84\%) of the fraction
of single-component sources in the  FIRST survey as reported by Cress
et al.~\cite{Cress} due to brighter flux density sample.
The fraction of
radio sources that are difficult to automatically identify is
about  15\% for a high angular resolution survey (such as FIRST).

The completeness of the RC catalog is close to unity in the central part of
the ``Cold'' for radio sources with flux densities \mbox{$S_{3.9\,GHz}>15$~mJy,}
and decreases toward the strip boundaries \cite{Soboleva}.
That is why to compare the parameters of radio sources, two complete samples
in the central part of the survey are considered. The first sample includes
the sources with elevation deviates from the
directional diagram of the telescope by less than
\mbox{$\Delta H\le |5\arcmin|$} and with flux densities that are greater than or
equal to $S_{3.9\,GHz}\ge 11$~mJy. The sample
covers an area of about 21\,$\square\degr$. The second sample includes the
sources with $\Delta H\le |10\arcmin|$
and $S_{3.9\,GHz}>29$~mJy (the total area is equal to about
41\,$\square\degr$).
The first (1S) and second (2S) samples contain a total of 130 and 117
objects, respectively. The two samples
partially overlap.

We computed the spectral index $\alpha_{1.4-4.85\,GHz}$ for all sources in
both samples and then subdivided these
sources into the following four groups:
\begin{itemize}
\item
inverse (I), $\alpha<-0.1$;
\item
flat (F), $0.1\le\alpha<0.5$;
\item
steep (S), $0.5\le\alpha<1$;
\item
ultrasteep (U), $\alpha\ge 1$.
\end{itemize}
A comparison of the two samples, with one of them being deeper than the
other in flux  density terms, shows that the number of
sources with inverse
and flat spectra in the frequency interval 1.4--4.85\,GHz slightly increases
with decreasing flux  density, i.e.,
the fraction of such sources is equal to  37\% and 28\% in the first and
second samples, respectively. The number
of sources with steep and ultrasteep spectra increases---the fraction of
such sources is equal to 63\% and 72\%
in the 1S and 2S samples, respectively. The distribution of the spectral
indices  $\alpha_{1.4-4.85\,GHz}$ of the
sources in the 2S sample is shifted toward steeper indices with respect to
the corresponding distribution for the
1S sample.

Sources with flat and inverse spectra in the 1S and 2S samples proved to be
more compact in terms of angular sizes
compared to sources with steep and ultrasteep spectra.

For sources identified in the  VLSS, TXS, NVSS, and GB6 catalogs additional
flux  density information is available allowing
the behavior of their radio spectra to be compared in the frequency
interval \mbox{74--4850\,MHz.} We computed
the two-frequency spectral indices $\alpha_{74-365\;MHz}$,
$\alpha_{365\;MHz-1.4\;GHz}$, and $\alpha_{1.4-4.85\;GHz}$
and fitted the spectra to a parabolic function. We could identify eight
groups of spectra and found the spectra of
most of the sources (about 60\%) to be flat or steep ones in the frequency
interval 74--365\;MHz and to remain
or become steep at high frequencies. About 10\% of the sources have S-shaped
spectra (the fractions of FSF and SFS
objects  are equal to about 3\% and 7\%, respectively). There are few (less
than 3\%) sources with a transition
from a flat to a steep or inverse spectrum. Radio sources with inverse
spectrum in one of the portions of the
frequency interval considered proved to be compact in terms of angular size.

We paid special attention to the morphological classification of radio
sources from the subsequent
optical identification point of view. We classify radio sources in
accordance with a morphological scheme based on the
comparison of the structure of the radio source and the position of the
optical object. We classified 39\% of the
objects as belonging to the point type (C core); 40\% of the sources as
double sources (D---double-lobe,
DC---double-core-lobe, DD---double-double), and about 20\% of all objects
as triple, multicomponent, and other types
of sources.

For the RC radio sources identified with  SDSS, USNO, and 2MASS catalogs
we found optical identifications,
which we will report in a separate paper.

\begin{acknowledgments}
We are grateful to E.~K.~Majorova for providing the computed
beam pattern of RATAN-600 radio telescope.

This work was supported by the Russian Foundation for Basic Research
(grant  \No 06-07-08062).
\end{acknowledgments}

\end{document}